\documentclass[prb,twocolumn,superscriptaddress,showpacs]{revtex4-1}

\usepackage{amsmath,amssymb}
\usepackage[dvipdfmx]{graphicx}
\usepackage{bm}
\usepackage{bbm}
\usepackage[hypertex,colorlinks,bookmarks=true,citecolor=blue,linkcolor=black,urlcolor=magenta]{hyperref}
\usepackage{color}
\usepackage{srcltx}
\usepackage{times,txfonts}

\allowdisplaybreaks

\begin{document}

\title{Nonreciprocal thermal and thermoelectric transport of electrons in noncentrosymmetric crystals
}
\date{\today}
\author{Ryota Nakai}
\affiliation{RIKEN Center for Emergent Matter Science (CEMS), Wako, Saitama 351-0198, Japan}
\email{rnakai@riken.jp}
\author{Naoto Nagaosa}
\affiliation{RIKEN Center for Emergent Matter Science (CEMS), Wako, Saitama 351-0198, Japan}
\affiliation{Department of Applied Physics, The University of Tokyo, Bunkyo, Tokyo 113-8656, Japan}

\begin{abstract}
 Nonreciprocal transport phenomena indicate that the forward and backward flows differ, and are attributed to broken inversion symmetry.
 In this paper, we study the nonreciprocity of the thermal and thermoelectric transport of electronic systems resulting from inversion-symmetry-broken crystal structures.
 The nonlinear electric, thermoelectric, and thermal conductivities are derived up to the second order in an electric field and a temperature gradient by using the Boltzmann equation with the relaxation time approximation.
 All the second-order conductivities appearing in this paper are described by two functions and their derivatives, and these are related to each other in the same way that linear conductivities are e.g.~via the Wiedemann-Franz law.
 We found that non-vanishing thermal-transport coefficients in the zero-temperature limit appear in nonlinear conductivities, which dominate the thermal transport at a sufficiently low temperature.
 The nonlinear conductivities and possible observable quantities are estimated in a $1H$ monolayer of the transition metal dichalcogenides MoS$_2$ and a polar semiconductor BiTeX(X=I,Br). 
 \end{abstract}

\maketitle


\section{Introduction}
\label{sec:introduction}

A system with broken inversion symmetry favors a certain direction of a larger flow than the opposite direction.
This property is known as nonreciprocity or rectification\cite{tokura18}.
A prominent example of electronic realization of the nonreciprocal transport occurs in a heterojunction of the p- and n-type semiconductors, where the geometry of two different materials breaks inversion symmetry.
Regulating the carrier motion by making a directed current is the key to device applications.
In a single crystalline system, the crystal structure is the source of inversion-symmetry breaking.
In noncentrosymmetric crystals, the nonreciprocity of the electronic charge transport requires either breaking of time-reversal symmetry or the presence of the electronic correlation and dissipation\cite{morimoto18}.

In addition to the electronic charge transport, managing a thermal transport is becoming growing issue since the heat is generated and fades ubiquitously in electronic devices\cite{giazotto06, roberts11}.
Unlike the electric current, the heat current is carried by any elementary excitations in materials.
The nonreciprocal thermal transport has been studied for each carrier, such as phonons\cite{terraneo02,li04,chang06,li12,torrent18} and photons\cite{otey10,roberts11}.
However, in this paper, we focus only on the thermal transport carried by the electrons in crystals.
Under an AC electric field $\bm{E}=\text{Re}[\bm{\mathcal{E}}e^{i\omega t}]$ and a gradient of temperature $-\bm{\nabla} T$, the linear electric and heat response is formulated in the matrix form as
\begin{align}
 \begin{pmatrix}
  \bm{j}\\
  \bm{j}^T
 \end{pmatrix}
 =\text{Re}
 \Bigg[
 \begin{pmatrix}
  L_{11} &
  L_{12}\\
  L_{21}&
  L_{22}
 \end{pmatrix}
 \begin{pmatrix}
  \bm{\mathcal{E}}
  e^{i\omega t}\\
  -
  \bm{\nabla} T/T
 \end{pmatrix}
 \Bigg].
 \label{eq:introlinearconductivities}
\end{align}
In the zero-frequency limit ($\omega=0$), the amount of charge and heat carried by electrons is closely connected through the Wiedemann-Franz law, which states that the ratio of two diagonal coefficients gives the constant $L_{22}/T^2L_{11}=\pi^2/3e^2\equiv \mathcal{L}$, independent of materials
(we use the natural unit $c=\hbar=k_\text{B}=1$ throughout this paper.)
This constant is referred to as the Lorentz number.
The other two of the off-diagonal linear conductivities are related by the Onsager's reciprocity as $L_{12}=L_{21}$.
{Moreover, the thermoelectric conductivity and the electric conductivity are related via the Mott formula representing the Seebeck coefficient $S$ as $L_{12}/T=L_{11}S=-e\mathcal{L}TdL_{11}/d\mu$.

Nonreciprocity is resulting from a nonlinear effect of the external field, where a strong external field drives the system far from equilibrium.
Extending the transport coefficients to the second order in the external fields, the electric and heat currents are shown in the matrix form as
\begin{align}
 \begin{pmatrix}
  \bm{j} \\
  \bm{j}^T
 \end{pmatrix}
 =
 \text{Re}
 \Bigg[
 \begin{pmatrix}
  L_{111}^{0} & L_{111}^{2\omega} & L_{112}^{\omega} & L_{122} & L_{13} \\
  L_{211}^{0} & L_{211}^{2\omega} & L_{212}^{\omega} & L_{222} & L_{23}
 \end{pmatrix}
 \begin{pmatrix}
  \bm{\mathcal{E}}\otimes
  \bm{\mathcal{E}}^{\ast}\\
  \bm{\mathcal{E}}\otimes
  \bm{\mathcal{E}}
  e^{2i\omega t}\\
  \bm{\mathcal{E}}e^{i\omega t}\otimes(-\bm{\nabla}T)/T\\
  \bm{\nabla} T/T\otimes\bm{\nabla} T/T\\
  (\bm{\nabla}\otimes\bm{\nabla} T)/T
 \end{pmatrix}
 \Bigg].
 \label{eq:conductivities_nonlinear}
\end{align}
Notice that we do not consider the oscillating temperature gradient, which is always assumed to be time-independent.
Here, each coefficient tensor in (\ref{eq:conductivities_nonlinear}) is specified by subscript indices in the same manner as those in (\ref{eq:introlinearconductivities}), indicating that the first index is 1(2) when the current is electric (heat) and the remaining two indices are 11, 22, and 12 when the applied external field is the square of an electric field, the square of a temperature gradient, and the combination of them, respectively.
When the second index is 3, the applied field is the square derivative of the temperature.
The nonlinear electric and heat currents proportional to the square of the electric field are decomposed into a part flowing constantly in one direction and the other part oscillating with the frequency $2\omega$.
Nonlinear conductivities for these two parts are denoted by superscripts $0$ and $2\omega$, respectively.
In addition, each coefficient is a 3-rank tensor having three spatial indices, that is, the electric current in (\ref{eq:conductivities_nonlinear}) is given by $j_a=\text{Re}[(L_{111}^0)_{abc}\mathcal{E}_b\mathcal{E}_c^\ast+\cdots]$ and so on.
Repeated indices in each term imply summation over the spatial index $a=x,y\,(,z)$.
The coefficients in (\ref{eq:conductivities_nonlinear}), or more generally, conductivity tensors for the even number of the external fields identically vanish when inversion symmetry is present, since the current, the electric field, and the temperature gradient are polar vectors.
In this paper, we address the coefficients in (\ref{eq:conductivities_nonlinear}).

From the perspective of application, exploring the nonreciprocal thermal transport provides a possible route toward functional thermal-transport phenomena such as electronic refrigeration\cite{giazotto06} and thermoelectric energy conversion\cite{sanchez13,cimmelli14}.
It has been shown that the nonlinear thermoelectric transport modifies the thermoelectric conversion efficiency besides the contribution from the thermoelectric figure of merit\cite{cimmelli14}.
So far, the nonreciprocal thermal and thermoelectric transport have been studied for electrons in nanojunctions\cite{segal05,kuo10,sanchez04,lopez13} and heterostructures\cite{fornieri14, martinez15}.
Inversion symmetry in such systems is broken by spatially asymmetric coupling or structures.
In this paper, we study the nonreciprocity of the thermal and thermoelectric transport of the electrons in two and three dimensions resulting from noncentrosymmetry of the crystal structure.
In combination with Zeeman coupling breaking time-reversal symmetry, the electric, thermal and thermoelectric nonreciprocal transports are realized.

This paper is organized as follows.
First, we briefly summarize main results of this paper in section \ref{sec:mainresults}.
In section \ref{sec:general}, the nonequilibrium distribution function under an AC electric field and a temperature gradient is derived by solving the Boltzmann equation with the relaxation time approximation, and the linear and the second-order electric, thermal, and thermoelectric conductivities are formulated.
A property unique to the nonlinear thermal and thermoelectric conductivity is that they contain temperature-independent terms while the linear terms vanish in the zero-temperature limit.
In section \ref{sec:zerot}, we explain the appearance of these zero-temperature conductivity from the viewpoint of nonequilibrium variation of a heat and an electric charge due to the external fields.
The nonlinear thermal and thermoelectric conductivities are estimated for the transition metal dichalcogenides MoS$_2$ in section \ref{sec:tmd}, and for a polar semiconductor BiTeX(X=I,Br) in section \ref{sec:bitex}.
Finally, we summarize our results in section \ref{sec:conclusion}.

\section{Main results}
\label{sec:mainresults}

Let us first present our main results so that readers are not confused with a number of physical quantities appearing in this paper.
Our main findings are that
(i) the nonlinear conductivities are described by two functions (\ref{eq:c_nonlinearextrinsic}) and (\ref{eq:d_berrycurvaturedipole}),
(ii) proportionality and derivative relations hold between the nonlinear conductivities, and
(iii) the nonlinear thermal and thermoelectric conductivities remain finite in the zero-temperature limit, although the corresponding linear conductivities vanish in the same limit due to vanishing thermal excitations.

We discuss two contributions to the nonlinear conductivities.
One is due to the nonlinear (proportional to the square of the relaxation time $\tau^2$) distribution function, described by a function
\begin{align}
 C_{abc}(\mu)
 =
 -
 \int_k
 v_av_bv_c
 \Theta(\mu-\epsilon_k),
 \label{eq:c_nonlinearextrinsic}
\end{align}
where $\bm{v}$ is the velocity, $\int_k$ is the abbreviation for the integral over the Brillouin zone and the summation over the band index $\sum_n\int d^dk/(2\pi)^d$, and $\Theta(x)$ is the step function given by 1 for $x>0$ and 0 for $x<0$.
The other is resulting from the combination of the intrinsic Berry curvature
$
\bm{\Omega}(\bm{k})
=
i
\langle
\partial u/\partial \bm{k}|
\times
|\partial u/\partial \bm{k}
\rangle
$
and the linear ($\propto \tau$) nonequilibrium distribution function, described by 
\begin{align}
 D_{ab}(\mu)
 =
 \int_k
 \frac{\partial \Omega_b}{\partial k_a}
 \Theta(\mu-\epsilon_k).
 \label{eq:d_berrycurvaturedipole}
\end{align}
(\ref{eq:d_berrycurvaturedipole}) is the zero-temperature part of the Berry curvature dipole\cite{sodemann15} which measures the dipole moment of the Berry curvature in the momentum space, and can be nonzero even when time-reversal symmetry is present, that is, when the integral of the Berry curvature itself over the Brillouin zone vanishes.

The leading-order term in the low-temperature (Sommerfeld) expansion of each nonlinear conductivity is independent of the temperature and is given by
\begin{align}
 &(L_{111}^0)_{abc}
 =
 \frac{e^3}{2(1+i\omega \tau)}
 \left(
 \frac{\tau^2}{2}
 \frac{d^2 C_{abc}}{d\mu^2}
 +
 \tau\epsilon_{abd}D_{cd}
 \right)
 +
 O(T^2),\\
 &(L_{111}^{2\omega})_{abc}
 =
 \frac{e^3}{2(1+i\omega \tau)}
 \left(
 \frac{\tau^2}{2(1+2i\omega \tau)}
 \frac{d^2 C_{abc}}{d\mu^2}
 +
 \tau\epsilon_{abd}D_{cd}
 \right)
 +
 O(T^2),\\
 &\frac{(L_{122})_{abc}}{T^2}
 =
 \frac{\pi^2e\tau^2}{3}\frac{d^2 C_{abc}}{d\mu^2}
 +
 O(T^2),\\
 &(L_{211}^0)_{abc}
 =
 -
 \frac{e^2\tau^2}{2(1+i\omega \tau)}
 \frac{d C_{abc}}{d\mu}
 +
 O(T^2),\\
 &
 (L_{211}^{2\omega})_{abc}
 =
 -
 \frac{e^2\tau^2}{2(1+i\omega \tau)(1+2i\omega\tau)}
 \frac{d C_{abc}}{d\mu}
 +
 O(T^2),\\
 &\frac{(L_{222})_{abc}}{T^2}
 =
 -\frac{\pi^2\tau^2}{3}
 \frac{d C_{abc}}{d\mu}
 +
 O(T^2).
\end{align}

In the absence of the Berry curvature dipole, or in case of negligible contribution from it (e.g. a nonlinear longitudinal transport such as $abc=xxx$), proportionality and derivative relations hold between the leading-order terms as
\begin{align}
 \frac{(L_{122})_{abc}/T^2}{2(L_{111}^0+L_{111}^{2\omega})_{abc}}
 \simeq
 \frac{(L_{222})_{abc}/T^2}{(L_{211}^0+L_{211}^{2\omega})_{abc}}
 \to
 \mathcal{L}\,(\omega\to 0),
 \label{eq:relationofnoninearheatconductivity}
\end{align}
and
\begin{align}
 2(L_{111}^{0/2\omega})_{abc}
 &\simeq
 -e\frac{d (L_{211}^{0/2\omega})_{abc}}{d\mu},
 \label{eq:derivativerelationofnonlinearconductivity0}\\
 (L_{122})_{abc}
 &\simeq
 -e\frac{d (L_{222})_{abc}}{d\mu}.
 \label{eq:derivativerelationofnonlinearconductivity}
\end{align}
Notice that $\omega\to 0$ limit of $L_{111}^0+L_{111}^{2\omega}$ and $L_{211}^0+L_{211}^{2\omega}$ appearing in the denominators of (\ref{eq:relationofnoninearheatconductivity}) is the DC conductivities.

\section{Nonlinear transport coefficients}
\label{sec:general}

In this section,
a nonequilibrium distribution function up to the second order in an electric field and a temperature gradient is derived from the Boltzmann equation.
We give formulae of the resulting linear and second-order conductivities, which lead to the leading-order terms shown in section \ref{sec:mainresults}.
Notice that there appear nonlinear conductivities for the product of the electric field and the temperature gradient $\bm{\mathcal{E}}e^{i\omega t}\otimes (-\bm{\nabla} T)/T$, and those for the square derivative of the temperature $(\bm{\nabla}\otimes\bm{\nabla} T)/T$, which are not shown in section \ref{sec:mainresults}, but are also contained in this section.

\subsection{Nonequilibrium distribution function}

We address transport properties in the nonlinear regime by the Boltzmann equation of the semiclassical dynamics of the electron, given by
\begin{align}
 \frac{\partial f}{\partial t}
 +
 \dot{r}_a
 \frac{\partial f}{\partial r_a}
 +
 \dot{k}_a
 \frac{\partial f}{\partial k_a}
 =
 -
 \frac{f-f_0}{\tau},
 \label{eq:boltzmannequation}
\end{align}
where $f_0=[1+e^{\beta (\epsilon_k-\mu)}]^{-1}$ is the Fermi-Dirac distribution function.
The semiclassical equations of motion in the presence of the Berry curvature, but in the absence of the magnetic field,
\begin{align}
 &\dot{\bm{r}}_a
 =
 \bm{v}
 -
 \dot{\bm{k}}
 \times
 \bm{\Omega}, \\
 &\dot{\bm{k}}
 =
 -
 e
 \bm{E}
 \label{eq:semiclassicalequationofmotion2}
\end{align}
describe the time evolution of the wave packet in the phase space, where
the velocity $\bm{v}$ and the Berry curvature $\bm{\Omega}$ are defined from the energy dispersion $\epsilon(k)$ and the Bloch function $u(k)$ by
$
v_a
=
\partial \epsilon/\partial k_a
$
and
$
\Omega_a
=
i
\epsilon_{abc}
\langle
\partial u/\partial k_b|
\partial u/\partial k_c
\rangle
$,
respectively.

Applying an AC electric field
$
 \bm{E}
 =
 \text{Re}[\bm{\mathcal{E}}e^{i\omega t}]
$,
the distribution function satisfies
\begin{align}
 \frac{\partial f}{\partial t}
 -
 eE_a
 \frac{\partial f}{\partial k_a}
 =
 -
 \frac{f-f_0}{\tau}.
 \label{eq:boltzmannequation_electricfield}
\end{align}
Up to the second order in the electric field, the nonequilibrium distribution function is written in the form of
\begin{align}
 f
 =
 f_0
 +
 \text{Re}
 \left[
 f_{E}^{\omega}
 +
 f_{E^2}^0
 +
 f_{E^2}^{2\omega}
 \right],
 \label{eq:distributionfunction_elecrticfield}
\end{align}
where the subscript of each term represents the order of the electric field and the superscript represents the frequency.
Substituting (\ref{eq:distributionfunction_elecrticfield}) into (\ref{eq:boltzmannequation_electricfield}), and making equations at each order in the electric field and the frequency, one obtains
\begin{align}
 f_{E}^{\omega}
 &=
 \frac{e\tau\mathcal{E}_ae^{i\omega t}}{1+i\omega\tau}
 \frac{\partial f_0}{\partial k_a}, \\
 f_{E^2}^0
 &=
 \frac{e^2\tau^2 \mathcal{E}_a \mathcal{E}_b^{\ast}}{2(1+i\omega\tau)}
 \frac{\partial^2 f_0}{\partial k_a\partial k_b},
 \label{eq:noneq_distribution_function_e20}\\
 f_{E^2}^{2\omega}
 &=
 \frac{e^2\tau^2 \mathcal{E}_a \mathcal{E}_b e^{2i\omega t}}{2(1+i\omega\tau)(1+2i\omega\tau)}
 \frac{\partial^2 f_0}{\partial k_a\partial k_b}.
 \label{eq:noneq_distribution_function_e22omega}
\end{align}
The zero-frequency term (\ref{eq:noneq_distribution_function_e20}) drives a current flowing in one direction, while a current caused by (\ref{eq:noneq_distribution_function_e22omega}) has oscillation with the frequency $2\omega$.

Applying a temperature gradient
$
 -\bm{\nabla} T
$,
the stationary distribution function satisfies
\begin{align}
 \dot{r}_a
 \frac{\partial f}{\partial r_a}
 =
 -
 \frac{f-f_0}{\tau},
\end{align}
where $f_0$, in this case, is the Fermi distribution function with an inhomogeneous temperature $\beta(\bm{r})=T^{-1}(\bm{r})$ representing local thermal equilibrium.
The solution up to the second order in the spatial derivative is
$
f
=
f_0
+
f_{\nabla T}
+
f_{\nabla^2 T}
+
f_{(\nabla T)^2}
$
where
\begin{align}
 f_{\nabla T}
 &=
 -
 \left(\nabla_a T\right)
 \tau v_a
 \frac{\partial f_0}{\partial T}, \\
 f_{\nabla^2 T}
 &=
 \left(\nabla_a\nabla_b T\right)
 \tau^2
 v_a v_b
 \frac{\partial f_0}{\partial T},\\
 f_{(\nabla T)^2}
 &=
 \left(\nabla_a T\right)
 \left(\nabla_b T\right)
 \tau^2
 v_a v_b
 \frac{\partial^2 f_0}{\partial T^2}.
\end{align}

In addition to the aforementioned terms, the second-order terms contain a mixed effect of the electric field and the temperature gradient given by
$
 f
 =
 \cdots +
 \text{Re}
 \left[
 f_{E\nabla T}^{\omega}
 \right]
$
where
\begin{align}
 f_{E\nabla T}^{\omega}
 =& 
 \frac{e\tau\mathcal{E}_ae^{i\omega t}
 \nabla_b T}{1+i\omega \tau}
 \left[
 \epsilon_{abc}
 \Omega_c
 \frac{\partial f_0}{\partial T}
 \right. \notag\\
 &\qquad\left.
 -
 \tau
 \frac{\partial v_b}{\partial k_a}
 \frac{\partial f_0}{\partial T} 
 -
 \left(
 1+
 \frac{1}{1+i\omega \tau}
 \right)
 \tau v_b
 \frac{\partial^2 f_0}{\partial T\partial k_a}
 \right].
\end{align}

\subsection{Electric current}

First, we formulate the electric current density
\begin{align}
 \bm{j}
 =
 -e
 \int_k
 \dot{\bm{r}} f
 -
 \bm{\nabla}
 \times
 \bm{m}^\text{orb}
 \label{eq:electriccurrentdefinition}
\end{align}
up to the second order in the external field.
The first term is the usual definition of the electric charge current, and the second term is the magnetization current\cite{cooper97,xiao10}.
Here we define the zero-temperature limit of the $\tau$-dependent DC electric conductivity $\sigma$ and the same limit of the intrinsic anomalous Hall conductivity $\sigma^\Omega$ by
\begin{align}
 \sigma_{ab}(\epsilon)
 &=
 e^2\tau
 \int_k
 v_av_b\delta(\epsilon-\epsilon_k),
 \label{eq:zeroT_linearextrinsic}\\
 \sigma_{ab}^\Omega(\epsilon)
 &=
 -e^2
 \epsilon_{abc}
 \int_k
 \Omega_c
 \Theta(\epsilon-\epsilon_k),
 \label{zeroT_linearintrinsic}
\end{align}
respectively.
As a consequence of the Wiedemann-Franz law, the Onsager's reciprocity, and the Mott formula, all the linear conductivities defined in (\ref{eq:introlinearconductivities}) are described by these two functions.
The orbital magnetization in the thermal equilibrium is closely related to the intrinsic anomalous Hall conductivity (\ref{zeroT_linearintrinsic}), since the circulating electric current due to the intrinsic Berry-phase effect is the source of the orbital magnetization.
The expression of the orbital magnetization can be derived by considering a confinement potential and estimating an electric current flowing along the edge due to the intrinsic effect\cite{xiao10}, and is given by
\begin{align}
 m_a^{\text{orb}}
 =
 -e^{-1}
 \epsilon_{abc}
 \int d\epsilon
 \sigma_{bc}^\Omega(\epsilon)
 f_0(\epsilon).
 \label{eq:orbitalmagnetization}
\end{align}
In a two-dimensional electronic system, the orbital magnetization points the out-of-plane direction.
Here we notice that although we have listed in (\ref{eq:conductivities_nonlinear}) the nonlinear conductivities resulting purely from the nonequilibrium distribution function  ($\propto \tau^2$) and those from the combination of the distribution and the intrinsic Berry curvature effect ($\propto \tau\bm{\Omega}$), it is possible that a purely intrinsic nonlinear conductivity that is independent of the relaxation time appears.
It has been reported that the intrinsic nonlinear conductivity is related to the orbital toroidal moment\cite{gao18} in a similar way as the intrinsic anomalous Hall conductivity is related to the orbital magnetic moment.
However, in this paper, we do not discuss the purely intrinsic contribution to the nonlinear conductivity, since it is independent of $\tau$ and is small compared with the terms mentioned when $\tau$ is large. 
 
Let us first review the AC electric conductivity defined by
\begin{align}
 j_a
 =
 \text{Re}
 \left[
 \left(
 L_{11}
 \right)_{ab}
 \mathcal{E}_b
 e^{i\omega t}
 +
 \left(
 L_{111}^{0}
 \right)_{abc}
 \mathcal{E}_b
 \mathcal{E}_c^{\ast}
 \right.&\notag\\
 \left.
 +
 \left(
 L_{111}^{2\omega}
 \right)_{abc}
 \mathcal{E}_b
 \mathcal{E}_c
 e^{2i\omega t}
 \right]&,
\end{align}
where 
\begin{align}
 &(L_{11})_{ab}
 =
 -
 \int d\epsilon\,
 \left(
 \frac{\sigma_{ab}}{1+i\omega\tau}
 +
 \sigma^\Omega_{ab}
 \right)
 \frac{\partial f_0}{\partial \epsilon}, 
 \label{eq:electricconductivity_linear}
\end{align}
$L_{111}^0=I_{111}^0+J_{111}$, and $L_{111}^{2\omega}=I_{111}^{2\omega}+J_{111}$ with
\begin{align}
 &(I_{111}^0)_{abc}
 =
 -
 \frac{e^3\tau^2}{4(1+i\omega\tau)}
 \int d\epsilon\,
 C_{abc}
 \frac{\partial^3 f_0}{\partial \epsilon^3},
 \label{eq:electricconductivity_nonlinear_zero}\\
 &(I_{111}^{2\omega})_{abc}
 =
 -\frac{e^3\tau^2}{4(1+i\omega\tau)(1+2i\omega\tau)}
 \int d\epsilon\,
 C_{abc}
 \frac{\partial^3 f_0}{\partial \epsilon^3},
 \label{eq:electricconductivity_nonlinear_2omega}\\
 &(J_{111})_{abc}
 =
 -
 \frac{e^3\tau\epsilon_{abd}}{2(1+i\omega\tau)}
 \int d\epsilon\,
 D_{cd}
 \frac{\partial f_0}{\partial \epsilon}.
 \label{eq:electricconductivity_nonlinear_berryphase}
\end{align}
Notice that in the above expressions, $C_{abc}$ and $D_{ab}$ in integrals are functions of $\epsilon$, not of the chemical potential $\mu$.

Next, the electric current induced by a spatial inhomogeneity of the local temperature profile $T(\bm{r})$ defines thermoelectric conductivities by
\begin{align}
 j_a
 =
 (L_{12})_{ab}
 \left(
 -
 \nabla_b T
 \right)/T
 +
 (L_{122})_{abc}
 (\nabla_b T)
 (\nabla_c T)/T^2\notag\\
 +
 (L_{13})_{abc}
 (\nabla_b\nabla_c T)/T
 ,
\end{align}
where 
\begin{align}
 &\frac{(L_{12})_{ab}}{T}
 =
 -e^{-1}
 \frac{\partial}{\partial T}
 \int d\epsilon\,
 \left(
 \sigma_{ab}
 +
 \sigma^\Omega_{ab}
 \right)
 f_0,
 \label{eq:thermoelectricbydt_linear}\\
 &\frac{(L_{122})_{abc}}{T^2}
 =
 -e\tau^2
 \frac{\partial^2}{\partial T^2}
 \int d\epsilon\,
 C_{abc}
 \frac{\partial f_0}{\partial \epsilon},
 \label{eq:thermoelectricbydt_nonlinear}\\
 &\frac{(L_{13})_{abc}}{T}
 =
 -e\tau^2
 \frac{\partial}{\partial T}
 \int d\epsilon\,
 C_{abc}
 \frac{\partial f_0}{\partial \epsilon}.
 \label{eq:thermoelectricbydt_nonlinear2}
\end{align}
Notice that the magnetization current resulting from the second term in (\ref{eq:electriccurrentdefinition}) contributes only to the linear conductivity (\ref{eq:thermoelectricbydt_linear}) since the magnetization is defined in local thermal equilibrium where higher-order corrections due to the external field are not considered.
An intriguing thing to be noticed is that $L_{122}/T^2$ can be finite in the zero-temperature limit $T\to 0$, while the other conductivities  $L_{12}/T$ and $L_{13}/T$ vanish linearly as the temperature goes down to zero.
This can be seen from the number of the temperature derivative since the Sommerfeld expansion contains even orders of the temperature.
The zero-temperature conductivity will be examined in section \ref{sec:zerot}.
Notice that, in a strict sense, the presence of a temperature gradient requires temperature to be nonzero.
However, we mean finiteness of the zero-temperature limit of a transport coefficient by the fact that, at a sufficiently low temperature, the transport coefficient has finite contributions independent of the temperature.

In the presence of both the electric field and the temperature gradient, there appear combined terms as
\begin{align}
 j_a
 =
 \text{Re}
 \left[
 \left(
 L_{112}^{\omega}
 \right)_{abc}
 \mathcal{E}_b
 e^{i\omega t}
 \left(
 -\nabla_c T
 \right)/T
 \right],
\end{align}
where $L_{112}^{\omega}=I_{112}^{\omega}+J_{112}^{\omega}$ with
\begin{align}
 &\frac{(I_{112}^{\omega})_{abc}}{T}
 =
 \frac{e^2\tau^2}{1+i\omega\tau}
 \frac{\partial}{\partial T}
 \int d\epsilon\,
 \left(
 \frac{1}{2}
 +
 \frac{1}{1+i\omega\tau}
 \right)
 C_{abc}
 \frac{\partial^2f_0}{\partial \epsilon^2},
 \label{eq:electricbyedt_nonlinear}\\
 &\frac{(J_{112}^\omega)_{abc}}{T}
 =
 -e^2\tau
 \frac{\partial}{\partial T}
 \int d\epsilon\,
 \left(
 \frac{\epsilon_{bcd}D_{ad}}{1+i\omega\tau}
 +
 \epsilon_{abd}D_{cd}
 \right)
 f_0.
 \label{eq:electricbyedt_nonlinear2}
\end{align}
Both conductivities (\ref{eq:electricbyedt_nonlinear}) and (\ref{eq:electricbyedt_nonlinear2}) vanish at the zero temperature.

\subsection{Heat current}

The heat current of the wave packet in the presence of the Berry phase is given by\cite{xiao06,xiao10}
\begin{align}
 \bm{j}^T
 =
 \int_k
 (\epsilon_k-\mu)
 \dot{\bm{r}}
 f
 +
 \bm{E}
 \times
 \bm{m}^{\text{orb}}.
\end{align}
The first term is the usual definition of the heat current, and the second term is the correction due to the Berry curvature.

The heat current induced by the AC electric field is written as
\begin{align}
 j^T_a
 =
 \text{Re}
 \left[
 \left(
 L_{21}
 \right)_{ab}
 \mathcal{E}_b
 e^{i\omega t}
 +
 \left(
 L_{211}^{0}
 \right)_{abc}
 \mathcal{E}_b
 \mathcal{E}_c^{\ast}
 \right. \notag\\
 \left.
 +
 \left(
 L_{211}^{2\omega}
 \right)_{abc}
 \mathcal{E}_b
 \mathcal{E}_c
 e^{2i\omega t}
 \right],
 \label{eq:heatcurrentbyelectricfield}
\end{align}
where 
\begin{align}
 &(L_{21})_{ab}
 =
 e^{-1}
 \int d\epsilon\,
 (\epsilon-\mu)\left(
 \frac{\sigma_{ab}}{1+i\omega\tau}
 +
 \sigma^\Omega_{ab}
 \right)
 \frac{\partial f_0}{\partial \epsilon},
 \label{eq:thermoelectricbye_linear}
\end{align}
$L_{211}^0=I_{211}^0+J_{211}$, and $L_{211}^{2\omega}=I_{211}^{2\omega}+J_{211}$ with
\begin{align}
 &(I_{211}^0)_{abc}
 =
 \frac{e^2\tau^2}{4(1+i\omega\tau)}
 \int d\epsilon\,
 (\epsilon-\mu)
 C_{abc}
 \frac{\partial^3 f_0}{\partial \epsilon^3},
 \label{eq:thermoelectricbye_nonlinear_zero}\\
 &(I_{211}^{2\omega})_{abc}
 =
 \frac{e^2\tau^2}{4(1+i\omega\tau)(1+2i\omega\tau)}
 \int d\epsilon\,
 (\epsilon-\mu)
 C_{abc}
 \frac{\partial^3 f_0}{\partial \epsilon^3},
 \label{eq:thermoelectricbye_nonlinear_2omega}\\
 &
 (J_{211})_{abc}
 =
 -
 \frac{e^2\tau\epsilon_{abd}}{2(1+i\omega\tau)}
 \int d\epsilon\,
 (\epsilon-\mu)
 D_{cd}
 \frac{\partial f_0}{\partial \epsilon}.
 \label{eq:thermoelectricbye_nonlinear_berryphase}
 \end{align}

The heat current induced by the spatial inhomogeneity of the temperature is written as
\begin{align}
 j^T_a
 =
 (L_{22})_{ab}
 \left(
 -\nabla_b T
 \right)/T
 +
 (L_{222})_{abc}
 (\nabla_b T)
 (\nabla_c T)/T^2\notag\\
 +
 (L_{23})_{abc}
 (\nabla_b\nabla_c T)/T,
\end{align}
which gives thermal conductivities by
\begin{align}
 &\frac{(L_{22})_{ab}}{T}
 =
 \frac{\partial}{\partial T}
 \int d\epsilon\,
 (\epsilon-\mu)
 \sigma_{ab}
 f_0,
 \label{eq:thermaldt_linear}\\
 &\frac{(L_{222})_{abc}}{T^2}
 =
 -\tau^2
 \frac{\partial^2}{\partial T^2}
 \int d\epsilon\, 
 (\epsilon-\mu)
 \frac{dC_{abc}}{d\epsilon}
 f_0,
 \label{eq:thermalddt_nonlinear2}\\
 &\frac{(L_{23})_{abc}}{T}
 =
 -\tau^2
 \frac{\partial}{\partial T}
 \int d\epsilon\, 
 (\epsilon-\mu)
 \frac{dC_{abc}}{d\epsilon}
 f_0.
 \label{eq:thermaldtdt_nonlinear}
\end{align}

Finally, combined terms of the electric field and the temperature gradient are defined by
\begin{align}
 j^T_a
 =
 \text{Re}
 \left[
 \left(
 L_{212}^{\omega}
 \right)_{abc}
 \mathcal{E}_b
 e^{i\omega t}
 \left(
 -
 \nabla_c T
 \right)/T
 \right],
\end{align}
where $L_{212}^{\omega}=I_{212}^{\omega}+J_{212}^{\omega}$ with
\begin{align}
 \frac{(I_{212}^{\omega})_{abc}}{T}
 =&
 \frac{e\tau^2}{1+i\omega\tau}
 \frac{\partial}{\partial T}
 \int d\epsilon\,
 \frac{dC_{abc}}{d\epsilon} \notag\\
 &\times
 \left(
 -\frac{1}{2}
 f_0
 +
 \left(
 \frac{1}{2}
 +
 \frac{1}{1+i\omega\tau}
 \right)
 (\epsilon-\mu)
 \frac{\partial f_0}{\partial \epsilon}
 \right),
 \label{eq:heatbyedt_nonlinear}\\
 \frac{(J_{212}^{\omega})_{abc}}{T}
 =&
 e\tau
 \frac{\partial}{\partial T}
 \int d\epsilon\,
 (\epsilon-\mu)
 \left(
 \frac{\epsilon_{bcd}D_{ad}}{1+i\omega\tau}
 +
 \epsilon_{abd}D_{cd}
 \right)
 f_0.
 \label{eq:heatbyedt_nonlinear2}
\end{align}
The low-temperature expansion of the nonlinear conductivities in this and the preceding subsections results in the leading-order terms shown in section \ref{sec:mainresults}.
The entire expression of the leading-order terms including those proportional to $T$ and $T^2$ is shown in Appendix \ref{sec:leadingorder}.

\subsection{Symmetry}

Symmetry property of the nonlinear conductivities can be read off from that of the two functions $C_{abc}$ and $D_{ab}$, since the nonlinear conductivities are described by them.
In this subsection, we consider how the conductivities are affected by the spatial-inversion and time-reversal transformations.
The results are summarized in Table \ref{table:symmetryproperty}.

In the presence of inversion symmetry, the velocity and the Berry curvature satisfy $v_a(n,-\bm{k})=-v_a(n,\bm{k})$ and $\Omega_a(n,-\bm{k})=\Omega_a(n,\bm{k})$, respectively, where $n$ is the band index.
In the presence of time-reversal symmetry, the velocity and the Berry curvature satisfy $v_a(n,-\bm{k})=-v_a(\bar{n},\bm{k})$ and $\Omega_a(n,-\bm{k})=-\Omega_a(\bar{n},\bm{k})$, respectively, where $\bar{n}$ is the band index that is the counterpart of the time-reversal pair of $n$.
Thus, regarding the linear response, the intrinsic anomalous Hall conductivity $\sigma^\Omega$ vanishes in the presence of time-reversal symmetry.

The function $C_{abc}$ changes its sign under either time-reversal or inversion transformation, that is, the presence of either symmetry forces a part of the nonlinear conductivities described by $C_{abc}$ to be identically zero.
On the other hand, the Berry curvature dipole $D_{ab}$ vanishes in the presence of inversion symmetry, since the presence of inversion and time-reversal symmetry imposes $(\Omega_b/k_a)(n,-\bm{k})=-(\Omega_b/k_a)(n,\bm{k})$ and $(\Omega_b/k_a)(n,-\bm{k})=(\Omega_b/k_a)(\bar{n},\bm{k})$, respectively.
More detailed crystallographic property of the Berry curvature dipole can be found in \onlinecite{sodemann15}.
\begin{table}[t]
 \begin{tabular}{cc||cccc}
  I & TR & $\sigma$ & $\sigma^\Omega$ & $C$ & $D$ \\
  \hline
  - & - & - & - & - & -\\
  - & + & - & 0 & 0 & -\\
  + & - & - & - & 0 & 0\\
  + & + & - & 0 & 0 & 0\\
 \end{tabular}
 \caption{Symmetry property of the functions representing the linear and the second-order conductivities under inversion and time reversal. The left two columns indicate cases of the presence (+) or the absence (-) of inversion (I) and time-reversal (TR) symmetry. The right four columns indicate that each tensor vanishes (0) or is not restricted (-) by corresponding symmetries.
 \label{table:symmetryproperty}}
\end{table}

\section{Thermal transport at zero temperature}
\label{sec:zerot}

As was shown in the previous section,
some of the nonlinear thermal and thermoelectric conductivities are finite in the zero-temperature limit.
In the absence of the Berry curvature dipole, they are described by a single function $C_{abc}(\mu)$, and thus proportionality relations (\ref{eq:relationofnoninearheatconductivity}) hold between them.
In this section, these properties are examined in connection with the relations between the \textit{linear} conductivities, that is, the Wiedemann-Franz law and the Mott formula.
Throughout this section, we consider negligible Berry curvature dipole and the limit of the vanishing frequency and temperature.

\subsection{Nonlinear heat current and dissipation}
The proportionality relation
\begin{align}
 \frac{(L_{222})_{abc}/T^2}{(L_{211}^0+L_{211}^{2\omega})_{abc}}
 \to
 \mathcal{L}\,(\omega\to 0)
\end{align}
between the zero-temperature limit of the nonlinear conductivities for the heat current is explained by the heat generated by external fields.
Let us consider the equation of the electronic energy in the presence of the external fields.
In general, a static equation of the energy is obtained by multiplying $\epsilon_k-\mu$ with the Boltzmann equation (\ref{eq:boltzmannequation}) and integrating over the momentum to give
\begin{align}
 \langle\epsilon_k-\mu\rangle
 =
 -
 \tau
 \bm{\nabla}\cdot
 \langle \bm{v} (\epsilon_k-\mu)\rangle
 +
 \tau
 \bm{E}
 \cdot
 \langle -e\bm{v}\rangle,
 \label{eq:energydynamics}
\end{align}
where $\langle\cdots\rangle=\int_k\cdots f$.
The lhs of (\ref{eq:energydynamics}) is the excess kinetic energy density of the electrons measured from the local-equilibrium value, and is proportional to the dissipation rate of the kinetic energy by relaxation processes.
The first term of the excess kinetic energy represents the divergence of the heat current, which appears when the heat current depends on the temperature.
The second term represents the Joule heat generated by the acceleration of the electrons by an electric field.
There appears the Lorentz number when we estimate the ratio of two energies in the linear regime as
\begin{align}
 \frac{-
 \bm{\nabla}\cdot
 \langle \bm{v} (\epsilon_k-\mu)\rangle}
 {\bm{E}
 \cdot
 \langle -e\bm{v}\rangle}
 =
 \frac{-
 \bm{\nabla}\cdot(L_{22}(-\bm{\nabla}T/T))}
 {\bm{E}\cdot
 L_{11}\bm{E}}
 \to
 \mathcal{L}
 \frac{|\bm{\nabla}T|^2}{|\bm{E}|^2}
 \label{eq:ratioofjouleheatanddiv}
\end{align}
in the zero-temperature limit,
since the linear electric and thermal conductivities are related by the Wiedemann-Franz law.

With these equations in mind, the nonlinear heat current induced by an electric field can be rewritten as
\begin{align}
 j_a^T
 &=
 \lim_{T\to 0}
 \int_k
 v_a
 \left[
 \tau\bm{E}\cdot
 (-e \bm{v})
 \right]
 f_E^{\omega\to 0}.
 \label{eq:heatcurrentbye_nonlinear_zero2}
\end{align}
This expression is analogous to the linear thermoelectric current $j^T_a=(L_{21})_{ab}E_b=\int_k v_a(\epsilon_k-\mu)f_E^{\omega\to 0}$, where the excitation energy $\epsilon_k-\mu$ is replaced by the Joule heat $\tau\bm{E}\cdot(-e \bm{v})$ appearing in (\ref{eq:energydynamics}).
In the similar way, the nonlinear heat current induced by the square of a temperature gradient is rewritten as
\begin{align}
 j^T_a
 &=
 \lim_{T\to 0}
 \int_k
 v_a
 \left[
 -\tau
 \bm{v}
 (\epsilon_k-\mu)
 \cdot
 \bm{\nabla}
 \right]
 f_{\nabla T},
 \label{eq:heatcurrentbye_nonlinear_zero3}
\end{align}
which, in comparison with the linear thermoelectric current representing the transport of the excitation energy, can be regarded as a transport of the divergence of the heat current, since
$-
\tau
\bm{\nabla}\cdot
\langle \bm{v} (\epsilon_k-\mu)\rangle
=
\int_k
[
-\tau
\bm{v}
(\epsilon_k-\mu)
\cdot
\bm{\nabla}
]
f
$.

The expressions (\ref{eq:heatcurrentbye_nonlinear_zero2}) and (\ref{eq:heatcurrentbye_nonlinear_zero3}) explain intuitively the reason why a heat current flows in the zero-temperature limit because the zero-temperature part of the second-order conductivities represents the transport of the Joule heat and the divergence of the heat current, respectively, not of the thermal excitation energy.
This also explains the reason why the zero-temperature nonlinear conductivities obey the proportionality relation because the Joule heat and the divergence of the heat current are related by the Lorentz number in the linear regime, as shown in (\ref{eq:ratioofjouleheatanddiv}) 
(see Appendix~\ref{sec:linearnonlinearproportionality} for a bit more rigorous argument regarding the relation between the linear and nonlinear conductivities.)

\subsection{Nonlinear electric current and charge density}

The proportionality relation
\begin{align}
 \frac{(L_{122})_{abc}/T^2}{2(L_{111}^0+L_{111}^{2\omega})_{abc}}
 \to
 \mathcal{L}\,(\omega\to 0)
\end{align}
of the electric current can also be explained by the linear conductivities.
The equation of the electric charge density is obtained by integrating the Boltzmann equation (\ref{eq:boltzmannequation}) as
\begin{align}
 \langle -e \rangle
 &=
 \tau
 \left[
 -e
 \bm{E}
 \cdot
 \frac{\partial \langle -e\bm{v}\rangle}{\partial \mu}
 -\bm{\nabla}T
 \cdot
 \frac{\partial \langle -e\bm{v}\rangle}{\partial T}
 \right],
 \label{eq:equationofelectriccharge}
\end{align}
where the gradient of the chemical potential is identified with the electric field $-\bm{\nabla}\mu=-e\bm{E}$.
The lhs of (\ref{eq:equationofelectriccharge}) is the electric charge density measured from its equilibrium value.
Provided external fields are applied in $x$ direction, the ratio of the electric charge density induced by a temperature gradient and that by an electric field is given in the linear regime by
\begin{align}
 \frac
 {-\bm{\nabla}T
 \cdot
 (\partial \langle -e\bm{v}\rangle/\partial T)}
 {-e\bm{E}
 \cdot
 (\partial \langle -e\bm{v}\rangle/\partial \mu)}
 &\to
 \mathcal{L}
 \frac{|\bm{\nabla}T|^2}{|\bm{E}|^2},
 \label{eq:ratio_chargedensity}
\end{align}
where the Mott formula $(L_{12})_{xx}/T=-e\mathcal{L}T(\partial (L_{11})_{xx}/\partial \mu)$ is used.

The property (\ref{eq:ratio_chargedensity}) explains intuitively the proportionality relation since the nonlinear electric current induced by an electric field and that by a temperature gradient are rewritten as
\begin{align}
 &2j_a
 =
 \int_k
 v_a
 \left[-e\tau \bm{E}\cdot(-e\bm{v})\frac{\partial}{\partial \mu}\right]
 f_{E}^{\omega\to 0},
 \label{eq:nonlinearelectric_bye_fe}\\
 &j_a
 =
 \lim_{T\to 0}\int_k
 v_a
 \left[-\tau \bm{\nabla} T\cdot (-e\bm{v})\frac{\partial}{\partial T}\right]
 f_{\nabla T},
 \label{eq:nonlinearelectric_bydt_fdt}
\end{align}
where the square brackets of (\ref{eq:nonlinearelectric_bye_fe}) and (\ref{eq:nonlinearelectric_bydt_fdt}) correspond to the electric charge density appearing in the first and the second term in the rhs of (\ref{eq:equationofelectriccharge}), respectively.
Therefore, the second-order electric current represents the transport of the electric charge raised by the external fields.

\section{Applications}

As applications of our theory, we estimate the thermal and thermoelectric coefficients of the electronic systems in noncentrosymmetric crystals.
Here, we focus only on the zero-temperature part of the nonlinear conductivity, since it dominates the transport at sufficiently low temperature.

\subsection{$1H$ monolayer of MoS$_2$}
\label{sec:tmd}

First, we consider the $1H$-type monolayer of the transition metal dichalcogenides, where the crystal has three-fold rotational symmetry and broken inversion symmetry.
The low-energy effective Hamiltonian accompanied with Zeeman coupling with an out-of-plane magnetic field is given by\cite{liu13}
\begin{align}
 H
 =
 \frac{\bm{k}^2}{2m}
 +
 \tau^z\lambda
 k_x\left(
 k_x^2-3k_y^2\right)
 -
 \Delta_\text{Z}\sigma^z
 -
 \Delta_\text{SO}\sigma^z\tau^z,
\end{align}
where $\sigma^z$ and $\tau^z$ are $\pm 1$ for each component of the spin and valley (K and K' points) degrees of freedom, respectively.
We assume that the Fermi level is above the bottoms of four energy bands, that is, $\mu>|\Delta_\text{Z}|+|\Delta_\text{SO}|$.
Perturbatively expanding by the parameter $\lambda$, the Fermi surface is modified up to the linear order in $\lambda$ as
\begin{align}
 k_\text{F}
 =
 \sqrt{2m\mu_{\sigma\tau}}
 -
 2m^2
 \mu_{\sigma\tau}
 \tau^z
 \lambda
 \cos
 3\theta
 \label{eq:tmd_fermisurface}
\end{align}
as a function of $\theta=\tan^{-1}(k_y/k_x)$,
where
$
\mu_{\sigma\tau}
=
\mu
+
\Delta_\text{Z}\sigma^z
+
\Delta_\text{SO}\sigma^z\tau^z
$.
The electric current has no nonreciprocal contribution within the relaxation-time approximation due to the cancellation by the spin and valley components\cite{wakatsuki17}.
Moreover, we cannot expect the nonlinear Berry-curvature effect described by the Berry curvature dipole,  since rotational symmetry cancels out the Berry curvature dipole in two dimensions.
However, the Berry curvature dipole in the $1T'$ and $1T_d$ structures of the transition metal dichalcogenides is being studied widely\cite{zhang18,zhang182,you18,xu18,ma18}.
Notice that the presence of the nonlinear electric conductivity $J_{111}$ due to the Berry curvature dipole in these materials implies the presence of another nonlinear conductivity $J_{212}^\omega$, since the leading-terms of these conductivities are described by $D_{ab}$.

\subsubsection{Nonlinear heat current}

The nonlinear DC heat current induced by an electric field at zero temperature is given by
\begin{align}
 \bm{j}^T
 =
 \frac{12 e^2\tau^2
 m\lambda\Delta_\text{Z}\Delta_\text{SO}}
 {\pi(1+\omega^2\tau^2)}
 \begin{pmatrix}
  |\mathcal{E}_x|^2 - |\mathcal{E}_y|^2 \\
  -\mathcal{E}_x\mathcal{E}_y^\ast-\mathcal{E}_x^\ast\mathcal{E}_y
 \end{pmatrix}.
 \label{eq:nonliearthermoelectric_tmd}
\end{align}
The DC heat current flows in one direction and can be distinguished from the other heat currents which oscillate and have vanishing total current when averaged over a period that is much longer than the inverse of the frequency.
From the proportionality relation (\ref{eq:relationofnoninearheatconductivity}), the nonlinear thermal conductivity is also nonvanishing in the zero-temperature limit, and is given by
\begin{align}
 \bm{j}^T
 =
 8\pi \tau^2
 m\lambda\Delta_\text{Z}\Delta_\text{SO}
 \begin{pmatrix}
  (\nabla_xT)^2 - (\nabla_yT)^2 \\
  -2(\nabla_xT)(\nabla_yT)
 \end{pmatrix}.
 \label{eq:nonliearthermal_tmd}
\end{align}
Regarding the derivative relation (\ref{eq:derivativerelationofnonlinearconductivity}), this result is consistent with the vanishing nonlinear electric current since the nonlinear heat current is independent of the chemical potential.

Even though the nonlinear heat currents (\ref{eq:nonliearthermoelectric_tmd}) and (\ref{eq:nonliearthermal_tmd}) dominate the external-field-driven heat current at a sufficiently low temperature, they are overwhelmed by the diffusion of a dissipated heat.
The external-field-driven heat current is proportional to the width of the sample and the diffusion of heat is proportional to the area of the sample.
The ratio of these two contributions is given by
\begin{align}
 \alpha
 =
 \frac{|\bm{j}^{T}|}{L\bm{j}\cdot\bm{E}}
 =
 \frac{6\tau m\lambda\Delta_\text{Z}\Delta_\text{SO}}{\mu L},
 \label{eq:ratio_nonlinearheat_dissipation}
\end{align}
where $\bm{j}^{T}$ is given in (\ref{eq:nonliearthermoelectric_tmd}), and $\bm{j}$ is the linear electric current.
For MoS$_2$, parameters estimated by \textit{ab initio} calculations $\hbar^2/2m=8.15$ eV\AA$^2$, $\lambda=-4.42$ eV\AA$^3$, $\Delta_\text{SO}=7.5$ meV\cite{liu13,saito15}, empirical parameters from $n\simeq 2m\mu/\pi\sim 10^{14}$ cm$^{-2}$, $\sigma_{xx}^{-1}\simeq\pi/2e^2\tau\mu=140$ $\Omega$\cite{wakatsuki17}, and an assumption $\Delta_\text{Z}=0.1$ meV give $\alpha\sim 10^{-6}$ for a sample with the linear dimension $L\sim 1$ $\mu$m.
This result indicates that $10^{-6}$ of a heat dissipated in a sample is transported in one direction due to the nonlinear effect.

\subsubsection{Nonlinear Ettingshausen and Hall effect}

Here, we consider a setup consisting of a transition metal dichalcogenides monolayer terminated by two electrodes.
If we apply an electric field in  $y$ direction between two electrodes, the nonlinear heat current flowing in $x$ direction will be safely separated from the diffusion of a dissipated heat flowing almost into the electrodes (in $y$ direction).
If the boundary perpendicular to $x$ direction is open, the nonlinear heat current generates a temperature difference and an electric potential difference between two boundaries to form a stationary state.
These nonlinear Ettingshausen effect and nonlinear Hall effect are estimated by requiring the transverse heat and electric current to vanish.
Then, we obtain
\begin{align}
 -\nabla_xT
 &=
 \frac{18e^2\tau m\lambda \Delta_\text{Z}\Delta_\text{SO}}
 {a\pi^2\mu T(1+\omega^2\tau^2)}
 |\mathcal{E}_y|^2, \\
 E_x
 &=
 \frac{6e\tau m\lambda\Delta_\text{Z}\Delta_\text{SO}}
 {a\mu^2}
 |\mathcal{E}_y|^2,
\end{align}
where $a=1-\pi^2/3\beta^2\mu^2$.
Considering a temperature far below the chemical potential ($k_\text{B}T\ll\mu$) and the frequency $\omega$ of the electric field much smaller than $\tau^{-1}$, we can replace $a$ and $1+\omega^2\tau^2$ by unity.
The coefficient of the nonlinear Ettingshausen effect of MoS$_2$ multiplied by the temperature is $T(-\nabla_xT/|\mathcal{E}_y|^2)=-3.5$ [K$^2$$\mu\text{m}/$V$^2$] and that of the nonlinear Hall effect is $E_x/|\mathcal{E}_y|^2=-3.4\times 10^{-7}$ [$\mu$m/V] from the same parameters used to estimate (\ref{eq:ratio_nonlinearheat_dissipation}).
Owing to the temperature-independent part of the nonlinear conductivity, the nonlinear Ettingshausen effect can be enhanced as the temperature becomes lower.

\subsection{Polar semiconductor BiTeX(X=I,Br)}
\label{sec:bitex}

Next, we consider a polar semiconductor BiTeX(X=I,Br), in which inversion symmetry is broken by the order of stacking layers\cite{ishizaka11}.
Inversion asymmetric crystal structure generates large bulk Rashba coupling.
Although the Berry curvature dipole of this material has been studied under a pressure\cite{facio18}, we focus only on a nonlinear transport independent of the Berry curvature, that is, those described by $C_{abc}$.
We consider the low-energy effective Hamiltonian under an in-plane magnetic field applied in $y$ direction, given by
\cite{ideue17}
\begin{align}
 H
 =
 \frac{k_z^2}{2m_{\parallel}}
 +
 \frac{k_x^2+k_y^2}{2m_{\perp}}
 +
 \lambda
 \left(k_x\sigma^y
 -
 k_y\sigma^x\right)
 -
 \Delta_\text{Z}\sigma^y.
 \label{eq:polarsemiconductorhamiltonian}
\end{align}
Here we assume that the Rashba parameter $\lambda$ is positive.
Results for a negative $\lambda$ can be obtained from the positive case by the spatial inversion, which inverts the sign of the second-order conductivities.

Let us first estimate two-dimensional conductivities carried by the electrons in the two-dimensional section of the whole Brillouin zone by fixing $\mu-k_z^2/2m_{\parallel}$.
The three-dimensional conductivity is evaluated by integrating the two-dimensional conductivity over $k_z$.
The shape of the two-dimensional Fermi surface changes as $k_z$ changes.
There are three regions in the space spanned by $\Delta_\text{Z}$ and $\mu-k_z^2/2m_{\parallel}$ distinguished by the Fermi surface topology and the helicity.
Three regions are
(I)
$
-|\Delta_\text{Z}|-m_{\perp}\lambda^2/2
<
\mu
<
|\Delta_\text{Z}|-m_{\perp}\lambda^2/2
$
having a single Fermi surface,
(II)
$
|\Delta_\text{Z}|-m_{\perp}\lambda^2/2
<
\mu
<
\Delta_\text{Z}^2/2m_{\perp}\lambda^2
$
having two Fermi surfaces with the same helicity, and
(III)
$
\Delta_\text{Z}^2/2m_{\perp}\lambda^2
<
\mu
$
having two Fermi surfaces with the opposite helicity\cite{ideue17}
(Fig.~\ref{fig:bitex_sigmaxxx}).
Notice that, on the boundary between the region II and III, the Fermi level lies at the charge neutral point of the linear dispersion, where the density of states of the inner Fermi surface vanishes and the semiclassical picture is no more applicable.

\subsubsection{Electric current}

\begin{figure}[t]
 \centering
 \includegraphics[width=72mm]{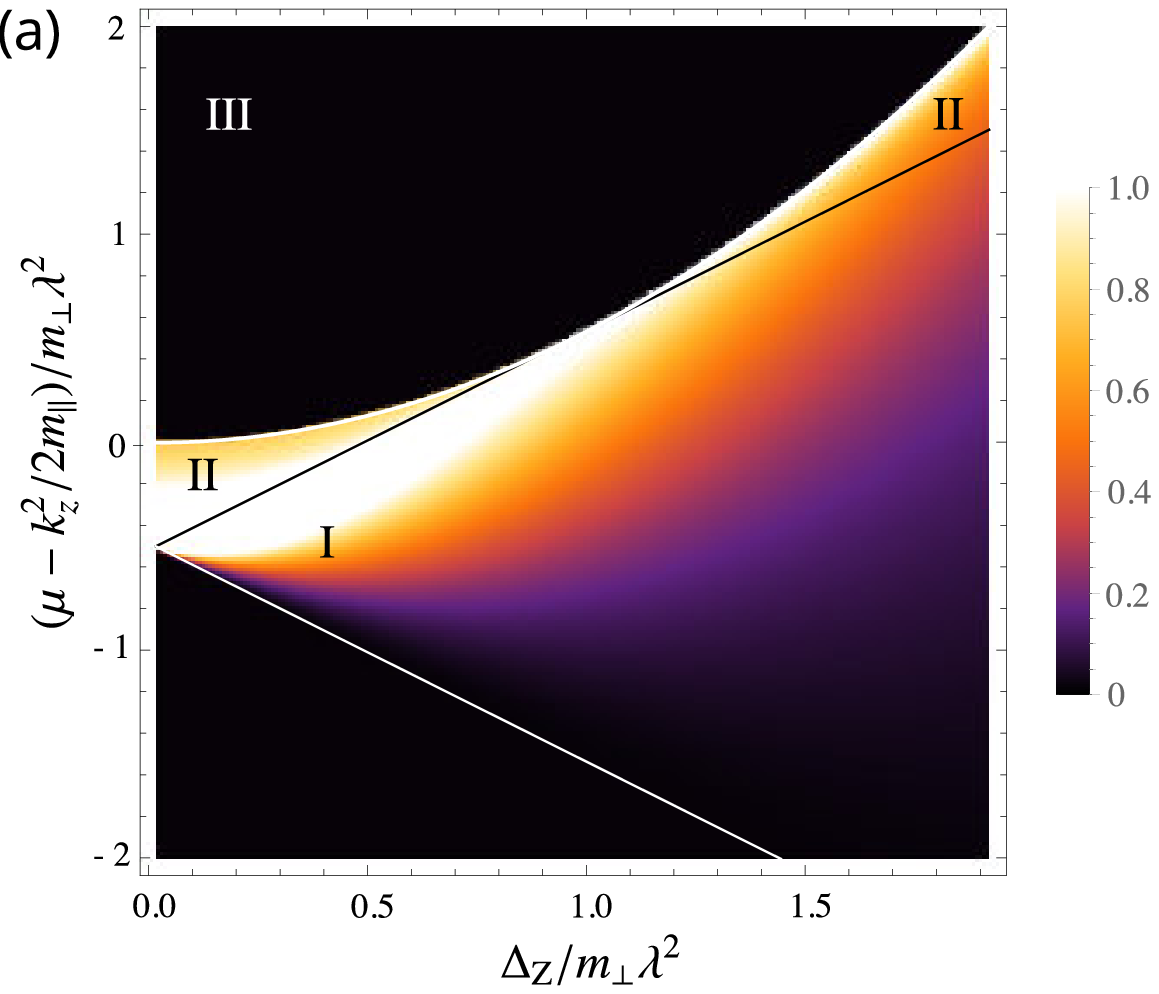}
 \vspace{3mm}\\
 \includegraphics[width=72mm]{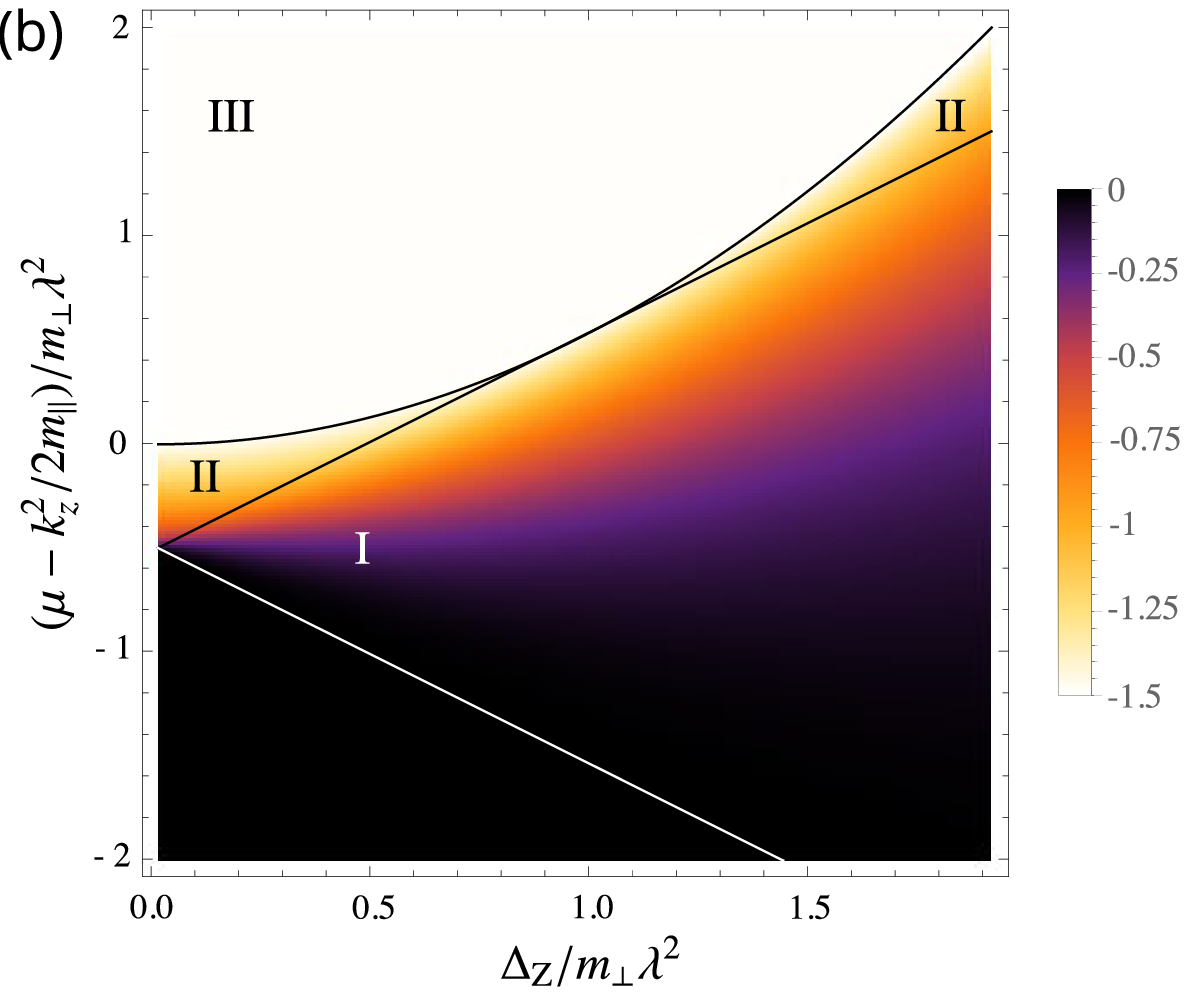}
 \caption{
 (a) The two-dimensional nonlinear conductivity of the electric current ($\propto (L_{111}^{0,\text{2D}})_{xxx}/\Delta_\text{Z}\propto (L_{122}^{\text{2D}})_{xxx}/\Delta_\text{Z}$) of BiTeX is shown as a function of the chemical potential and Zeeman coupling.
 (b) The two-dimensional nonlinear conductivity of the heat current ($\propto (L_{211}^{0,\text{2D}})_{xxx}/\Delta_\text{Z}\propto (L_{222}^{\text{2D}})_{xxx}/\Delta_\text{Z}$) in BiTeX is shown.
 \label{fig:bitex_sigmaxxx}}
\end{figure}

The electric conductivity up to the second order in the electric field has been reported in \onlinecite{ideue17}, and is given by
\begin{align}
 (L_{111}^{0,\text{2D}})_{xxx}
 =
 \frac{3e^3\tau^2}{16\pi(1+i\omega\tau)}
 \frac{\Delta_\text{Z}}{\sqrt{2m_{\perp}\mu'}}
 +
 O(T^2, \Delta_\text{Z}^2)
\end{align}
in the region II, where $\mu'=\mu+m_{\perp}\lambda^2/2$ is the chemical potential measured from the bottom of the energy band, and by
$
 (L_{111}^{0,\text{2D}})_{xxx}
 =
 O(T^2, \Delta_\text{Z}^2)
$
in the region III.
From (\ref{eq:relationofnoninearheatconductivity}), the nonlinear thermoelectric conductivity is obtained by $(L_{122}^\text{2D})_{xxx}/T^2=4\mathcal{L}(1+i\omega\tau)(L_{111}^{0,\text{2D}})_{xxx}$.
Numerical integration is performed to estimate the nonlinear electric conductivity $(L_{111}^{0,\text{2D}})_{xxx}/(e^3\tau^2\Delta_\text{Z}/4\pi m_{\perp}\lambda(1+i\omega\tau))$, which is shown in Fig.~\ref{fig:bitex_sigmaxxx} (a).

Let us analytically estimate the three-dimensional nonlinear electric conductivity.
A magnetic field 1T corresponds to $\Delta_\text{Z}/m_\perp\lambda^2\sim 6\times 10^{-3}$ for BiTeI and $2\times 10^{-2}$ for BiTeBr\cite{ideue17} by using parameters $m_\perp=0.15m_e$ ($m_e$ is the electron mass), $m_\parallel=5m_\perp$, $\lambda=3.85$ eV$\AA$ for BiTeI, and $\lambda=2.00$ eV$\AA$ for BiTeBr estimated by \textit{ab initio} calculation.
So, it is reasonable to evaluate the three-dimensional conductivities in the vanishing-magnetic-field limit, where the two-dimensional conductivity divided by Zeeman coupling $L_{111}^{0,\text{2D}}/\Delta_\text{Z}$ is finite while the conductivity itself vanishes\cite{ideue17}.
In this limit, it is possible to evade considering the contribution from the region I, where the analytical expression of the two-dimensional conductivities is hard to be obtained (see Fig.~\ref{fig:bitex_sigmaxxx}).
We obtain 
\begin{figure}[t]
 \includegraphics[width=84mm]{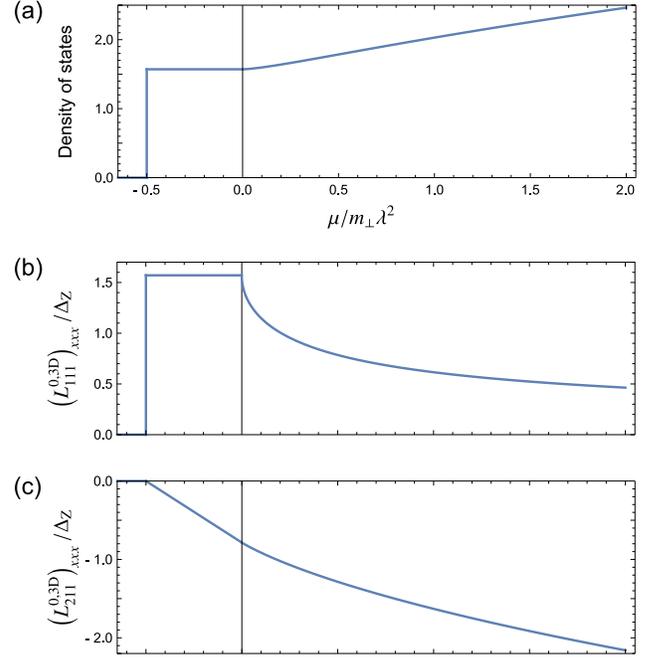}
 \caption{For BiTeX(X=I,Br), (a) the density of states in unit of $\sqrt{m_\perp^3 m_\parallel}\lambda/\pi^2$, (b) the three-dimensional nonlinear electric conductivity $(L_{111}^{0,\text{3D}})_{xxx}/\Delta_\text{Z}$ (proportional to the thermoelectric conductivity $(L_{122}^\text{3D})_{xxx}/\Delta_\text{Z}$) in unit of $3e^3\tau^2\sqrt{m_\parallel/m_\perp}/16\pi^2(1+i\omega\tau)$, and (c) the three-dimensional nonlinear thermoelectric conductivity $(L_{211}^{0,\text{3D}})_{xxx}/\Delta_\text{Z}$ (proportional to the thermal conductivity $(L_{222}^\text{3D})_{xxx}/\Delta_\text{Z}$) in unit of $3e^2\tau^2\lambda^2\sqrt{m_\perp m_\parallel}/8\pi^2(1+i\omega\tau)$ are plotted as functions of the chemical potential $\mu$ in unit of $m_\perp \lambda^2$. All quantities are estimated in the limit of the vanishing magnetic field ($\Delta_\text{Z}\to 0$).
 \label{fig:bitex_3ds}}
\end{figure}
\begin{align}
 (L_{111}^{0,\text{3D}})_{xxx}
 \simeq
 \left\{
 \begin{array}{l}
  \displaystyle
   \frac{3e^3\tau^2 \Delta_\text{Z}\sqrt{m_{\parallel}/m_{\perp}}}
   {32\pi(1+i\omega\tau)}\quad
   (-m_{\perp}\lambda^2/2<\mu<0)\\
  \displaystyle
   \frac{3e^3\tau^2 \Delta_\text{Z}\sqrt{m_{\parallel}/m_{\perp}}}
   {16\pi^2(1+i\omega\tau)}
   \cos^{-1}\!\!
   \sqrt{
   \frac{\mu}{\mu'}}\quad
   (\mu>0)
 \end{array}
 \right.
 \label{eq:bitex_3d2nd_electricconductivity}
\end{align}
after integrating over $k_z$ [Fig.~\ref{fig:bitex_3ds}(b)].

\subsubsection{Heat current}

In two dimensions, the thermoelectric conductivity of the heat current induced by the electric field is given by
\begin{align}
 (L_{211}^{0,\text{2D}})_{xxx}
 =
 \left\{
 \begin{array}{ll}
  \displaystyle
  -\frac{3e^2\tau^2\Delta_\text{Z}}{8\pi(1+i\omega\tau)}
   \frac{\sqrt{2m_{\perp}\mu'}}
   {m_{\perp}}
   +
   O(T^2, \Delta_\text{Z}^2)& (\text{II}) \\
  \displaystyle
  -\frac{3e^2\tau^2\Delta_\text{Z}}{8\pi(1+i\omega\tau)}\lambda
   +
   O(T^2, \Delta_\text{Z}^2)& (\text{III})
 \end{array}
 \right..
\end{align}
From (\ref{eq:relationofnoninearheatconductivity}), the nonlinear thermal conductivity is obtained by $(L_{222}^\text{2D})_{xxx}/T^2=2\mathcal{L}(1+i\omega\tau)(L_{211}^{0,\text{2D}})_{xxx}$.
In the entire region, numerical estimation of the nonlinear conductivity $(L_{211}^{0,\text{2D}})_{xxx}/(e^2\tau^2\lambda \Delta_\text{Z}/4\pi(1+i\omega\tau))$ is shown in Fig.~\ref{fig:bitex_sigmaxxx} (b).
Unlike the case of the electric current, the nonlinear heat current is finite even if the Fermi surface has opposite helicities (seen in the region III).

The three-dimensional nonlinear conductivity of the heat current is evaluated in the same manner as the electric current case, and is given by
\begin{align}
 (L_{211}^\text{0,3D})_{xxx}
 =
 &-\frac{3e^2\tau^2\Delta_\text{Z}\sqrt{m_{\parallel}/m_{\perp}}}
 {8\pi^2(1+i\omega\tau)}\notag\\
 &\times
 \left\{
 \begin{array}{l}
  \pi \mu'/2 \quad (-m_{\perp}\lambda^2/2<\mu<0)\\
  \sqrt{\mu(\mu'-\mu)}
   +
   \mu'
   \cos^{-1}\!\!
   \sqrt{\mu/\mu'}
   \quad (\mu>0)
 \end{array}
 \right..
\end{align}
The derivative relation (\ref{eq:derivativerelationofnonlinearconductivity}) indicates that the positive nonlinear electric conductivity (\ref{eq:bitex_3d2nd_electricconductivity}) leads to monotonically decreasing thermoelectric conductivity [see Fig.~\ref{fig:bitex_3ds}(c)].

\subsubsection{Nonlinear Seebeck effect}

The density of states is given by [Fig.~\ref{fig:bitex_3ds}(a)]
\begin{align}
 \frac{dn}{d\mu}
 =
 \frac{\sqrt{m_\perp^3m_\parallel}\lambda}{\pi^2}
 \times
 \left\{
 \begin{array}{l}
  \displaystyle
   \frac{\pi}{2}
   \,\, (-m_{\perp}\lambda^2/2<\mu<0)\\
  \displaystyle
    \cos^{-1}\!\!\sqrt{\frac{\mu}{\mu'}}
    +
    \sqrt{
    \frac{2\mu}{m_\perp\lambda^2}
    }\,\, (\mu>0)
 \end{array}
 \right..
 \label{eq:bitex_dos}
\end{align}
The second term of the density of state for $\mu>0$ converges to that for the quadratic energy band without Rashba coupling $dn/d\mu=\sqrt{2m_\parallel\mu}m_\perp/\pi^2$ in the limit $\lambda\to 0$. 
An empirical value $n=4\times 10^{19}$cm$^{-3}$ for BiTeI\cite{sakano13} corresponds to $\mu/m_\perp\lambda^2=-0.24$ ($\mu'=75$ meV), and $n=4\times 10^{18}$cm$^{-3}$ for BiTeBr corresponds to $\mu/m_\perp\lambda^2=-0.32$ ($\mu'=14$ meV).
Thus, it is legitimate to consider the case $\mu<0$, where conductivities have relatively simple forms.

We consider a setup consisting of a sample terminated by two heat baths at different temperature.
The ratio of the linear and nonlinear Seebeck coefficients is given by
\begin{align}
 \frac{S^\text{NL}}{S^\text{L}}
 =
 \frac{(L_{122})_{xxx}(\nabla_x T)^2/T^2}{(L_{12})_{xx}(-\nabla_x T)/T}
 =
 \frac{3\tau \Delta_\text{Z}}{2m_\perp \lambda}
 \frac{\nabla T}{T}.
\end{align}
When a magnetic field 1T and a temperature difference $\Delta T/T=0.1$ between both sides of a sample of the linear dimension 1 $\mu$m are applied, empirical values $g\sim 60$\cite{park13} and $(L_{11})_{xx}^{-1}=0.3$ m$\Omega$ cm for BiTeI\cite{ishizaka11} give $S^\text{NL}/S^\text{L}=4\times 10^{-5}$.

\section{Conclusion}
\label{sec:conclusion}

We studied nonreciprocal thermal and thermoelectric transport phenomena of the electrons resulting from an inversion-symmetry-broken energy band structure.
We derived the electric, thermoelectric, and thermal conductivities up to the second order in an electric field and a temperature gradient by solving the Boltzmann equation with the relaxation time approximation.
These second-order conductivities describe the nonreciprocal transport of the electric charge and the heat.

The nonlinear conductivities due to the nonequilibrium distribution function are nonvanishing when both inversion and time-reversal symmetries are broken, and those due to the combination of the nonequilibrium distribution function and intrinsic Berry curvature effect are nonvanishing when inversion symmetry is broken.
The nonlinear conductivities are related to each other by proportionality or derivative relations, which originate from the fact that all the nonlinear conductivities defined in this paper are formulated by two functions.
Moreover, the leading-order terms of the nonlinear thermal and thermoelectric conductivities in the low-temperature (Sommerfeld) expansion are independent of temperature, unlike the corresponding linear conductivities.
We explained the proportionality relation and the temperature-independence from the fact that the nonlinear electric and heat currents represent the transport of an electric-charge variation and a heat generated by external fields, respectively.

We have estimated the temperature-independent part of the nonlinear conductivities for the $1H$ monolayer of the transition metal dichalcogenides MoS$_2$ and a polar semiconductor BiTeX(X$=$I,Br) by using the low-energy effective Hamiltonian.
In the transition metal dichalcogenides monolayer under an out-of-plane magnetic field, we showed that the nonreciprocal heat current appears by applying an electric field or a temperature gradient.
We also showed that the nonreciprocal heat current can be separated by the other currents in a sample and that the nonlinear Ettingshausen effect can be a signature of the nonlinear heat current.
In BiTeX under an in-plane magnetic field, both nonlinear electric and heat currents occur in longitudinal direction. We estimated the nonlinear contribution to the Seebeck coefficient in comparison with the linear contribution.

\acknowledgements
This work was supported by Grant-in-Aid for Scientific Research Nos.~JP17K17604, JP24224009 and JP26103006 from the Ministry of Education, Culture, Sports, Science and Technology (MEXT), Japan.
R.N. was supported by RIKEN Special Postdoctoral Researcher Program.
N.N. was supported by the Impulsing Paradigm Change through Disruptive Technologies Program of Council for Science, Technology and Innovation (Cabinet Office, Government of Japan), and Core Research for Evolutionary Science and Technology
(CREST) No.~JPMJCR16F1.

\appendix

\section{The leading-order contributions}
\label{sec:leadingorder}
In this appendix, we present all the leading terms in the low-temperature expansion of the second-order conductivities studied in this paper.
There are 14 terms representing the nonlinear electric and heat currents (when a conductivity tensor $L_{ijk}$ is decomposed to $I_{ijk}$ and $J_{ijk}$, we count them distinctively as two.)
The nonlinear electric and heat currents are given by
\begin{widetext}
\begin{align}
 \begin{pmatrix}
  j_a \\ j_a^T
 \end{pmatrix}
 \simeq&
 \text{Re}\Bigg[
 -\frac{e^2}{2(1+i\omega \tau)}
 \left[
 \tau^2
 \begin{pmatrix}
  -eC''_{abc}/2\\
  C_{abc}'
 \end{pmatrix}
 +
 \tau
 \begin{pmatrix}
  -e
  \epsilon_{abd}D_{cd}\\
  \pi^2 T^2\epsilon_{abd}D_{cd}'/3
 \end{pmatrix}
 \right]
 \mathcal{E}_b
 \mathcal{E}^{\ast}_c \notag\\
 &\quad\,\,\,
 -\frac{e^2}{2(1+i\omega \tau)}
 \left[
 \frac{\tau^2}{1+2i\omega\tau}
 \begin{pmatrix}
  -eC''_{abc}/2\\
  C_{abc}'
 \end{pmatrix}
 +
 \tau
 \begin{pmatrix}
  -e
  \epsilon_{abd}D_{cd}\\
  \pi^2 T^2\epsilon_{abd}D_{cd}'/3
 \end{pmatrix}
 \right)
 \mathcal{E}_b
 \mathcal{E}_c
 e^{2i\omega t} \notag\\
 &\quad\,\,\,
 -\frac{\pi^2e T}{3(1+i\omega\tau)}
 \Bigg[
 \frac{\tau^2}{2}
 \begin{pmatrix}
  -e(1+2/(1+i\omega\tau))C_{abc}''' \\
  (3+4/(1+i\omega\tau))C_{abc}''
 \end{pmatrix} 
 -
 \tau
 \begin{pmatrix}
  -e \left(
  \epsilon_{bcd}D_{ad}'
  +
  (1+i\omega\tau)\epsilon_{abd}D_{cd}'
  \right) \\
  \epsilon_{bcd}D_{ad}
  +
  (1+i\omega\tau)\epsilon_{abd}D_{cd}
 \end{pmatrix}
 \Bigg]
 \mathcal{E}_b
 e^{i\omega t}
 (-\nabla_c T)
 \Bigg]\notag\\
 &-
 \frac{\pi^2\tau^2}{3}
 \begin{pmatrix}
  -eC_{abc}''\\
  C_{abc}'
 \end{pmatrix}
 \left(
 \nabla_b T\nabla_c T
 +
 T
 \nabla_b\nabla_cT
 \right),
 \end{align}
\end{widetext}
where the prime indicates the derivative with respect to the chemical potential.
Here we notice that subleading terms are not shown while some of them are of the same order of the leading term of the conductivity $J_{211}$.

\section{Proportionality relations of linear and nonlinear conductivities}
\label{sec:linearnonlinearproportionality}

Let us explain the proportionality relation between the second-order conductivities (\ref{eq:relationofnoninearheatconductivity}) from the perspective of the Wiedemann-Franz law for the linear conductivities.
By the Sommerfeld expansion of the integrand of the linear electric and thermal conductivities, we obtain
\begin{align}
 &(L_{11})_{ab}
 =
 -e^2\tau\int_k
 c_{ab}(\mu;n,\bm{k})
 +
 O(T^2),
 \label{eq:linearelectric_momentumsommerfeld}\\
 &\frac{(L_{22})_{ab}}{T}
 =
 -\frac{\pi^2\tau T}{3}\int_k
 c_{ab}(\mu;n,\bm{k})
 +
 O(T^3),
 \label{eq:linearheat_momentumsommerfeld}
\end{align}
where $c_{ab}(\epsilon;n,\bm{k})=v_av_b\delta(\epsilon-\epsilon_k)$ and $n$ is the band index.
This indicates that the same relation as the Wiedemann-Franz law holds for each momentum and the band index.
Also, the relation (\ref{eq:ratioofjouleheatanddiv}) between the Joule heat and the divergence of the heat current holds for each momentum and the band index in the linear regime, that is,
\begin{align}
 &\tau\bm{E}\cdot (-e\bm{v})f_E^{\omega\to 0}
 =
 e^2\tau^2E_aE_bc_{ab}(\mu;n,\bm{k})+O(T^2),\\
 &-\tau\bm{\nabla}\cdot
 \bm{v}(\epsilon-\mu)f_{\nabla T}
 =
 \frac{\pi^2\tau^2}{3}
 (\nabla_aT)(\nabla_bT)
 c_{ab}(\mu;n,\bm{k})+O(T^2).
\end{align}
Make sure that integrating over the momentum and summation of the band index turns the above equations back to (\ref{eq:ratioofjouleheatanddiv}).

Then, we rewrite the zero-temperature limit of the second-order heat currents (\ref{eq:heatcurrentbye_nonlinear_zero2}) and (\ref{eq:heatcurrentbye_nonlinear_zero3}) in terms of $c_{ab}$ as
\begin{align}
 &j^T_a
 =
 e^2\tau^2E_bE_c
 \int_k
 v_a
 c_{bc}(\mu;n,\bm{k})+O(T^2),\\
 &j_a^T
 =
 \frac{\pi^2\tau^2}{3}
 (\nabla_bT)(\nabla_cT)
 \int_k
 v_ac_{bc}(\mu;n,\bm{k})+O(T^2),
\end{align}
where the proportionality $(L_{222}/T^2)/(L_{211}^0+L_{211}^{2\omega})\to \mathcal{L}(\omega\to 0)$ between the second-order conductivities becomes obvious, and is reminiscent of the Wiedemann-Franz law of the linear conductivities.

In a similar way, the electric-charge variations induced by an electric field and a temperature gradient are related by the Mott formula and the relevant second-order electric currents (\ref{eq:nonlinearelectric_bye_fe}) and (\ref{eq:nonlinearelectric_bydt_fdt}) are written by
\begin{align}
 &2j_a
 =
 e^2\tau^2 E_bE_c
 \int_k
 (-e v_a)\frac{dc_{bc}(\mu;n,\bm{k})}{d\mu}
 +O(T^2),\\
 &j_a
 =
 \frac{\pi^2\tau^2}{3}
 (\nabla_bT)(\nabla_cT)
 \int_k
 (-e v_a)\frac{dc_{bc}(\mu;n,\bm{k})}{d\mu}
 +O(T^2),
\end{align}
where $dc_{bc}(\mu;n,\bm{k})/d\mu$ appears in describing the Mott formula in the linear response regime.

\end{document}